\documentclass[3p,times,twocolumn]{elsarticle}

\usepackage{ecrc}

\volume{00}

\firstpage{1}

\journalname{Nuclear and Particle Physics Proceedings}

\runauth{Rainer J.\ Fries}

\jid{nppp}

\jnltitlelogo{Nuclear and Particle Physics Proceedings}

\usepackage{amssymb,amsmath}
\usepackage{graphicx}

\usepackage[figuresright]{rotating}

\begin{document}

\begin{frontmatter}



\dochead{}

\title{A Novel Approach For Event-By-Event Early Gluon Fields}


\author{Rainer J.\ Fries}
\author{Steven Rose}
\address{Cyclotron Institute and Department of Physics and Astronomy, Texas A\&M University, College Station TX, USA}

\begin{abstract}
We report on efforts to construct an event generator that calculates the classical gluon field generated at early times
in high energy nuclear collisions. Existing approaches utilize numerical solutions of the Yang-Mills
equations after the collision. In contrast we employ the analytically known recursion relation in the forward light
cone. The few lowest orders are expected to lead to reliable results for times of up to the inverse saturation scale,
$\tau_0 \sim 1/Q_s$. In these proceedings we sketch some calculational details, and show some preliminary results.
\end{abstract}


\end{frontmatter}



The initial phase of collisions of heavy nuclei at very high energies is believed to be be described by
color glass condensate (CGC) as an effective theory \cite{Gelis:2010nm}. The initial wave functions and the collision system
just after the collision are dominated by classical gluon fields \cite{McLerran:1993ka,McLerran:1993ni,JMKMW:96,Kovchegov:1996ty,KoMLeWei:95}.
The density of gluons and their average transverse momentum in the initial wave functions is given by a new energy scale, 
the saturation scale $Q_s$. At times $\tau_0\sim 1/Q_s$ after the collision fast growing quantum fluctuations lead
to a demise of the purely classical picture and drive the system toward kinetic equilibrium \cite{Romatschke:2006nk,Dusling:2012ig,Berges:2013eia}. Eventually a quark gluon plasma close to local kinetic equilibrium emerges.

The boundary value problem for classical fields after the collision was first formulated in \cite{KoMLeWei:95}. Several
groups have since solved the problem numerically, see e.g.\ \cite{Krasnitz:1998ns,Krasnitz:2002mn,Lappi:2003bi,Fukushima:2011nq}. More recently the IP-Glasma model \cite{Schenke:2012wb,Schenke:2012fw} combined 
numerical solutions of the Yang Mills problem with an event-by-event sampling of initial charge densities 
according to the IP-Sat model \cite{Bartels:2002cj,Kowalski:2003hm} and a subsequent matching to viscous fluid dynamics \cite{Gale:2012rq}. 
In these proceedings we report on our efforts to build an even generator based on an alternative approach \cite{Rose:2016hgy}.
We use the recursive solutions to the Yang-Mills boundary value problem in the forward light cone discussed
in \cite{Fries:2006pv,Chen:2013ksa,Chen:2015wia}. For a gauge field $(x^+A,-x^-A,A_\perp^i)$ in light cone coordinates 
in Fock-Schwinger gauge we use a power series $A=\sum_n A_{(n)} \tau^n$, and similar for the transverse field $A_\perp^i$, $=1,2$ . The 
recursion relations solving the Yang-Mills equations read
\begin{align}
  A_{(n)} =& \frac{1}{n(n+2)} \sum_{k+l+m=n-2} \left[ D^i_{(k)}, \left[
  D^i_{(l)}, A_{(m)} \right] \right] ,  \nonumber \\
  A^i_{\perp(n)} =& \frac{1}{n^2}\left( \sum_{k+l=n-2}
  \left[ D^j_{(k)}, F^{ji}_{(l)} \right] \right. \label{eq:a_recursion2}
  \\
  &+ \left.
  ig \sum_{k+l+m=n-4} \left[ A_{(k)}, [ D^i_{(l)},A_{(m)} ] \right] \right)
  \, ,
  \nonumber
\end{align}
with initial conditions
\begin{align}
  A_{\perp(0)}^i  &= A_1^i  + A_2^i \, ,
  \label{eq:bc_boost4}\\
  A_{(0)} &= -\frac{ig}{2} \left[ A_1^i , A_2^i \right] \, ,
  \label{eq:bc_boost5}
\end{align}
where $A_1^i$ and $A_2^i$ ($i=1,2$) are the fields in the two nuclei before the collision.
Using the few lowest orders of the recursion should give reasonable results up to the time $\tau_0$.
The obvious advantage of our approach for an event-by-event calculation is that no differential equations 
need to be solved for the time evolution. The numerical effort simply goes into calculating coefficients of a power series
and is reasonably fast. On the other hand, the convergence of the power series is limited to early times.
In the preliminary results reported below we restrict ourselves to the second order. Systematic tests are under way
to see how far the recursion can be pushed while mainting numerical stability and low run times.

In the following we discuss some of the steps implemented so far. Event averages are known
analytically for a number of quantities that can be used to check the accuracy of the 
event-by-event Monte Carlo code. Of course the matching of analytic event-averaged results is only a necessary, 
not a sufficient, criterion. Eventually the code should be benchmarked against existing numerical 
solvers event-by-event. The three main steps in the code are
\begin{enumerate}
\item Sampling of the transverse color charge densities $\rho_k^a(x,y)$ for nuclei $k=1,2$ and $a=1,\ldots,N_c^2-1$.
\item Computation of the nuclear fields $A_1^i$, $A_2^i$ ($i=1,2$) from the nuclear charges, respectively.
\item Calculation of the field after the collision using the recursion relation.
\end{enumerate}

\begin{figure}[tb]
\includegraphics[width=\columnwidth]{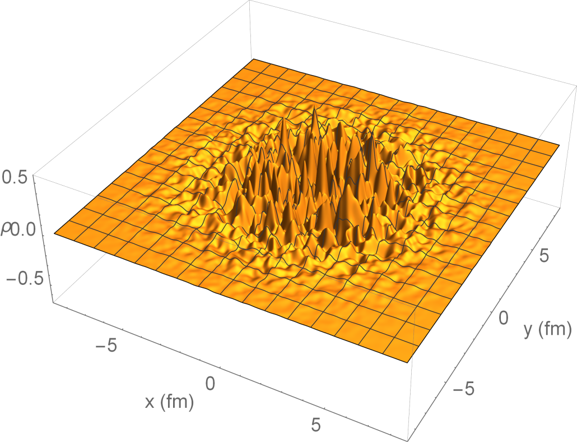}
\caption{\label{fig:charge} The color component $\rho^1$ of a sampled color charge for a nucleus of radius $R_A = 3$ fm (arbitrary units). The granularity is determined by the coarse graining scale $\lambda$.}
\end{figure}

\begin{figure}[tb]
\includegraphics[width=\columnwidth]{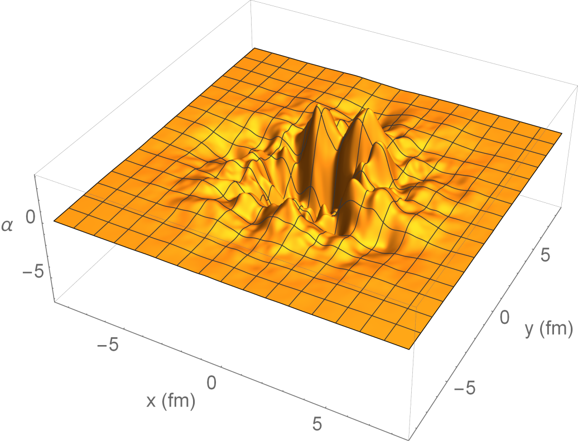}
\caption{\label{fig:alpha} The color component $\alpha^1$ of the covariant potential for the same event as shown in
Fig.\ \ref{fig:charge} (arbitrary units). Transverse structures for the covariant potential are dominated
by the infrared scale $m$.}
\end{figure}

\begin{figure}[tb]
\includegraphics[width=\columnwidth]{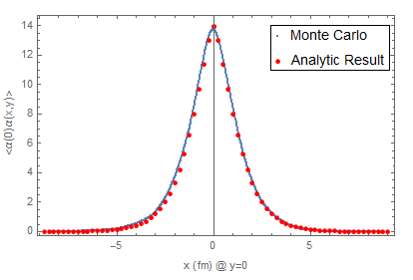}
\caption{\label{fig:correlation} The correlation function $\langle\alpha^a(\mathbf{r}_1) \alpha^a(\mathbf{r}_2)
\rangle$ (arbitrary units) calculated with $\mathbf{r}_1$ fixed at the center of the nucleus as a function
of $\mathbf{r}_2$ moving along the $x$-axis. The Monte-Carlo result using 7800 events and the analytic result
\cite{Chen:2015wia} are shown.}
\end{figure}

\begin{figure}[tb]
\includegraphics[width=\columnwidth]{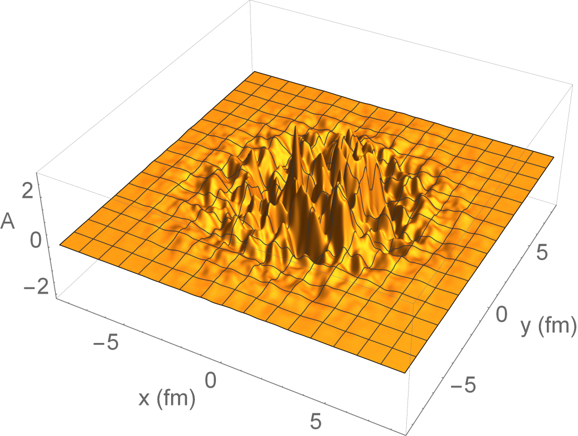}
\caption{\label{fig:A} The $x$-component of the gauge field, $A^{1,1}$ (single color) for the same event as shown
in Fig.\ \ref{fig:charge} (arbitrary units). The dependence on the infrared scale is suppressed after the gauge 
transformation.}
\end{figure}

{\bf Sampling.} The sampling assumes the Gaussian charge distributions of the McLerran-Venugopalan model \cite{McLerran:1993ka,McLerran:1993ni},
$\langle\rho^a_k\rangle=0$, $\langle \rho_k^a \rho_k^b\rangle \sim \delta^{ab} \mu_k(x,y)$ (see \cite{Chen:2015wia} for 
details). For the results shown here it is assumed that the functional dependence 
of the variance $\mu_k(x,y)$ of the charge distribution is given by the nuclear thickness function of the 
respective nucleus. More physical choices, including correlations from nucleons, or the IP-Sat model 
could be implemented. Fig.\ \ref{fig:charge} shows a typical example for one sampled color component of a nuclear charge
distribution.

{\bf Nuclear Fields.} To arrive at gauge fields in Fock-Schwinger gauge we use the standard approach shown
in \cite{JMKMW:96}. We solve the Yang-Mills equation for a single nucleus first in covariant gauge to obtain a covariant potential
$\alpha_k(x,y)$. Subsequently we apply a gauge transformation to the desired gauge. $\alpha_k(x,y)$ for each nucleus
is the solution to a Poisson equation. We solve the Poisson equation by applying a pre-tabulated coarse-grained Greens 
function
\begin{equation}
  G\left(\mathbf{r}\right) = \frac{1}{2\pi^2\lambda^2}\int d^2 z K_0\left( m \mathbf{z} \right) 
  e^{-\left(\mathbf{r}-\mathbf{z}\right)/\lambda^2}
\end{equation}
where $m$ is the infrared scale for the Poisson problem. The coarse graining scale $\lambda$ acts as a 
UV cutoff that can be conveniently chosen to rid the results of lattice artefacts, $1/Q_s \gg \lambda \gg a$ where
$a$ is the lattice constant. Fig.\ \ref{fig:alpha} shows one color component for a typical covariant potential.
Fig.\ {\ref{fig:correlation} shows one necessary test of the numerical implementation. In this case the correlation function 
$\langle\alpha^a(\mathbf{r}_1) \alpha^a(\mathbf{r}_2) \rangle$ is computed numerically over 7800 events
and compared to the analytic result. Both correlation functions agree very well.
The gauge transformation requires us to compute
\begin{equation}
  A^i_k  =-\frac{1}{ig} U_k \left(\mathbf{r} \right) \partial^i U_k^\dagger\left( \mathbf{r}\right)
\end{equation}
from the path-ordered exponential
\begin{equation}
  U_k\left(\mathbf{r}\right) = \mathcal{P} \exp\left[ ig\int dz^\mp \alpha_k\left(\mathbf{r},z^\mp \right) \right] \, .
\end{equation}
The integral over the longitudinal direction is realized by summing over a discrete set of uncorrelated covariant 
gauge fields $\alpha_k$ along $z^\mp$ with the same $\mu_k$.
Fig.\ \ref{fig:A} shows a typical example of one color component of the field $A^1$ in Fock-Schwinger gauge.

\begin{figure}[tb]
\includegraphics[width=\columnwidth]{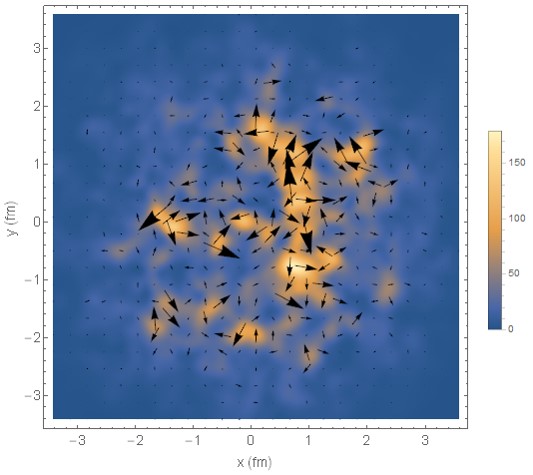}
\caption{\label{fig:flow2} The initial energy density $\varepsilon_0$ 
 (shading), with energy flow $T^{0i}$ overlayed (arrows) for a single head-on collision of $R_A=3$ fm nuclei
(arbitary units).
The space-time rapidity $\eta$ is taken to be very large so that effects of $\beta$ to the energy flow are of
the same order of magnitude as the hydro-like flow $\alpha$.}
\end{figure}

{\bf Collision.} 
The initial longitudinal chromo-electric and chromo-magnetic fields after the collision can be 
computed readily from commutators of fields of the two nuclei
\begin{equation}
  E_0 =  ig \delta^{ij} \left[ A_1^i,  A_2^j \right], \quad B_0 =  ig \epsilon^{ij} \left[ A_1^i,  A_2^j \right] \, .
\end{equation}
We will focus here on the energy momentum tensor of the gluon field after the collision. In Ref.\ \cite{Chen:2015wia}
its lowest order terms are given explicitly as commutators and covariant derivatives of the initial fields $E_0$ and
$B_0$. The energy density at $\tau=0$ is $\varepsilon_0 = (E_0^2+B_0^2)/2$ and the initial transverse energy flow is
\begin{equation}
  T^{0i} = \frac{\tau}{2}\left( \alpha_i \cosh\eta + \beta_i \sinh\eta \right) + \mathcal{O}(\tau^3)
\end{equation}
($i=1,2$) where $\eta$ is the space-time rapidity and the rapidity-even and rapidity-odd flow terms are
\begin{align}
  \alpha_i &= -\nabla^i \varepsilon_0 \, , \\
  \beta_i & =\epsilon^{ij} \left( [D^j,B_0]E_0 - [D^j,E_0]B_0\right) \, .
\end{align}
Fig.\ \ref{fig:flow2} shows an example for a central Pb+Pb event at large rapdity $\eta \gg 1$ with the transverse energy 
flow $T^{0i}$ superimposed on the initial energy density $\varepsilon_0$. The $\alpha$-flow follows gradients 
of the energy density, mimicking hydro-like behavior. In contrast the $\beta$-flow is determined by the 
underlying dynamics of the non-abelian gauge fields. It is, for example, responsible for the transport of angular
momentum toward mid-rapidity in non-central collisions \cite{Fries:2017}.

\begin{figure}[tb]
\includegraphics[width=\columnwidth]{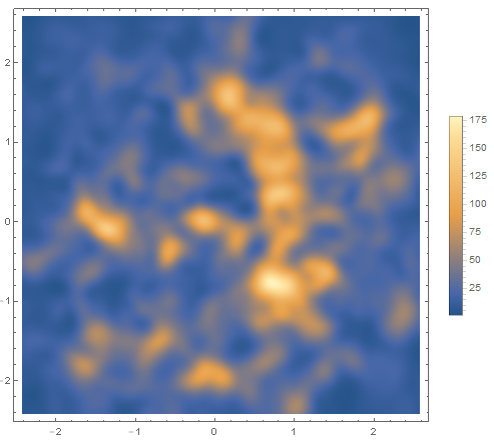}
\includegraphics[width=\columnwidth]{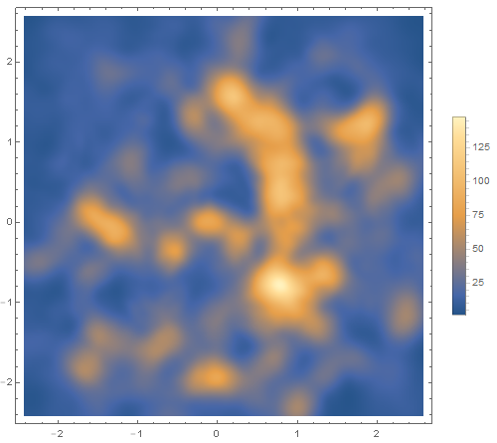}
\caption{\label{fig:time} Time evolution of the energy density $T^{00}$ in the transverse plane for a
head-on collision of $R_A=3$ fm nuclei at $\eta=0$ (arbitrary units). Upper Panel: $\tau=0$ fm/$c$. 
Lower Panel: $\tau=0.1$ fm/$c$. Note the change of scale indicated by the legends on the right.}
\end{figure}

Fig.\ \ref{fig:time} shows the time evolution of the energy density for a typical collision between $\tau=0$ and 
$\tau=0.1$ fm/$c$ using the recursion relation to second order. The obvious effect of the brief time evolution is
an expansion and diffusion of "hot spots" in the energy density through the build up of flow.

In summary, we have shown preliminary results from an event generator for early-time classical gluon fields that is
based on the recursive solution of the Yang Mills equations in the forward light cone. Obvious improvements 
can be made by (1) going to more realistic models of the average charge densities $\mu_k$, and by (2) pushing
to higher orders in the recursion relation. Work on both issues is underway. Current checks rely on comparisons
of event averages that are known analytically. Further checks of event-by-event results are desirable.

RJF would like to thank the organizers of Hard Probes 2016 for a wonderful conference. This work was supported 
by the US National Science Foundation under award no.\ 1516590 and award no.\ 1550221.


\nocite{*}
\bibliographystyle{elsarticle-num}
\bibliography{jos}



\end{document}